\documentclass[aps,prl,groupedaddress,twocolumn]{revtex4}

\usepackage{bbm}
\usepackage{mathrsfs}
\usepackage{amsmath}
\usepackage{amsfonts}
\usepackage[colorlinks=true,citecolor=blue,anchorcolor=blue]{hyperref}
\usepackage{graphicx,epstopdf}
\usepackage{subfigure}
\usepackage{epsfig}
\usepackage{dcolumn}
\usepackage{bm}
\usepackage{color}
\usepackage{natbib}
\usepackage{amssymb}
\usepackage{xcolor}
\usepackage{braket}

\begin{document}
\title{Nonlinear level attraction of cavity axion polariton in an antiferromagnetic topological insulator}

\author{Yang Xiao$^{1,\ast}$, Huaiqiang Wang$^{2,3,\ast}$, Dinghui Wang$^{2}$, Ruifeng Lu$^{4}$, Xiaohong Yan$^{5}$, Hong Guo$^{6}$, C. -M. Hu$^{7}$, Ke Xia$^{8}$, Haijun Zhang$^{2,9,\dagger}$ and Dingyu Xing$^{2,9}$ }
\address{$^1$College of Science, Nanjing University of Aeronautics and Astronautics, Nanjing 210016, China}
\address{$^2$National Laboratory of Solid State Microstructures and Physics School, Nanjing University, Nanjing 210093, China}
\address{$^3$Department of Physics, University of Zurich, Zurich 8057, Switzerland}
\address{$^4$Department of Applied Physics, Nanjing University of Science and Technology, Nanjing 210094, China}
\address{$^5$School of Material Science and Engineering, Jiangsu University, Zhenjiang, 212013, China}
\address{$^6$Department of Physics, McGill University, Montreal, Quebec H3A 2T8, Canada}
\address{$^7$Department of Physics and Astronomy, University of Manitoba, Winnipeg R3T 2N2, Canada}
\address{$^8$Beijing Computational Science Research Center, Beijing 100193, China}
\address{$^9$Collaborative Innovation Center of Advanced Microstructures, Nanjing University, Nanjing 210093, China}

\email{zhanghj@nju.edu.cn}

\begin{abstract}
\textbf{
Strong coupling between cavity photons and elementary excitations in condensed matters boosts the field of light-matter interaction and generates several exciting sub-fields, such as cavity optomechanics and cavity magnon polariton. Axion quasiparticles, emerging in topological insulators, were predicted to strongly couple with the light and generate the so-called axion polariton. Here, we demonstrate that there arises a gapless level attraction of cavity axion polariton in an antiferromagnetic topological insulator, which originates from the high-order interaction between axion and the odd-order resonance of a cavity. Such a novel coupling mechanism is essentially different from conventional level attractions with the linear-order interaction. Our results open up promising roads for exploring the axion polariton with cavity technologies. }
\end{abstract}


\maketitle

\begin{figure*}[t]
\centering
\includegraphics[angle=0,scale=0.8]{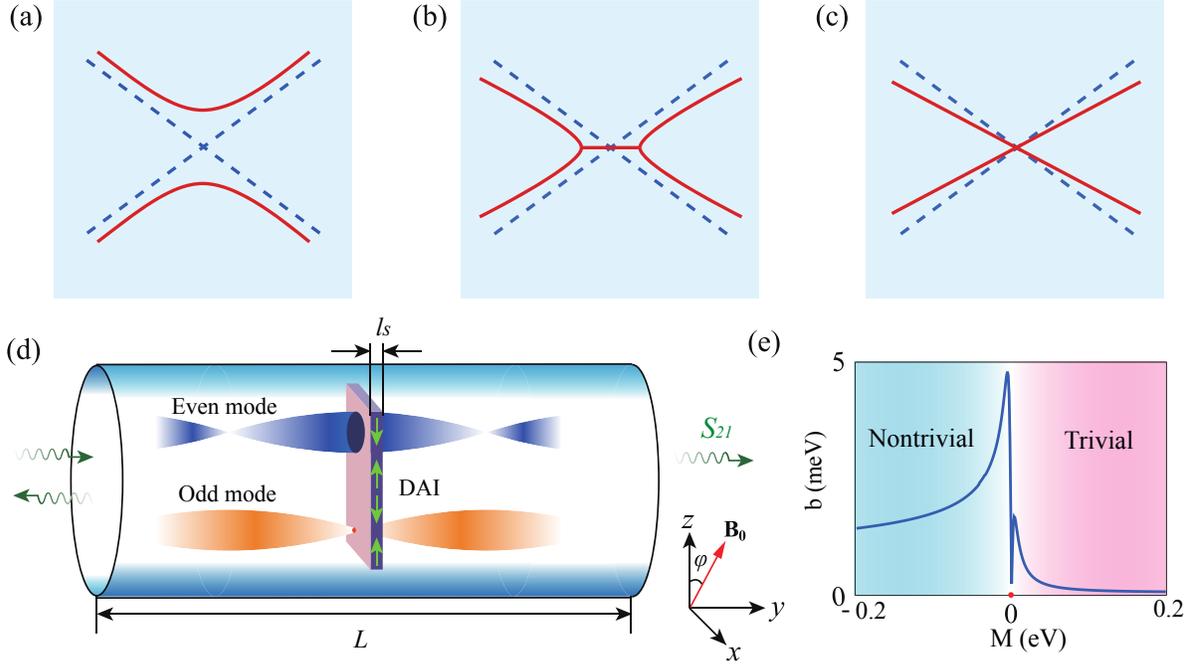}
\caption{  Conventional LR (\textbf{a}), conventional LA (\textbf{b}) and high-order LA (\textbf{c}) depicted with solid red lines. The dashed blue lines generally present two uncoupled modes. (\textbf{d}) Schematic of cavity geometry with a dynamical axion insulator (DAI) film. The photon propagates from the left port to the right port through the DAI film. Depending on the mode number, the cavity resonance mode can be the even mode or the odd mode. As for the odd mode, the electric field at the DAI is negligibly small, indicating that the photon-axion coupling would be dominated by the high-order interaction. But for the even mode, a large electric field can lead to a linear photon-axion coupling. The static magnetic field is applied with an angle $\varphi$ with respect to the $z$ axis and induces a coupling between the electric field $\mathbf{E}$ of photon and dynamical axion field (DAF) $\delta \theta$. (\textbf{e}) The coupling strength $b$ between axion mode and the photon as a function of the mass parameter $M$, distinguishing between topologically trivial ($M>0$) and nontrivial ($M<0$) DAI. } \label{fig1}
\end{figure*}

\begin{figure}[!ht]
\centering
\includegraphics[angle=0,scale=0.48]{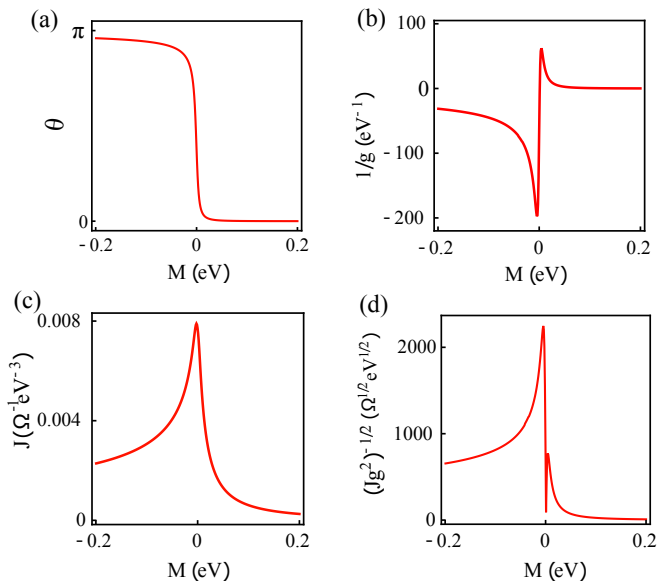}
\caption{(\textbf{a}) The axion angle, (\textbf{b}) coefficient $1/g$, (\textbf{c}) stiffness $J$, and (\textbf{d}) $1/(g\sqrt{J})$ as a function of the mass term $M$, where $M<0$ ($M>0$) corresponds to the topologically nontrivial (trivial) phase with(without) band inversion.} \label{fig2}
\end{figure}

\begin{figure*}[!ht]
\centering
\includegraphics[angle=0,scale=0.8]{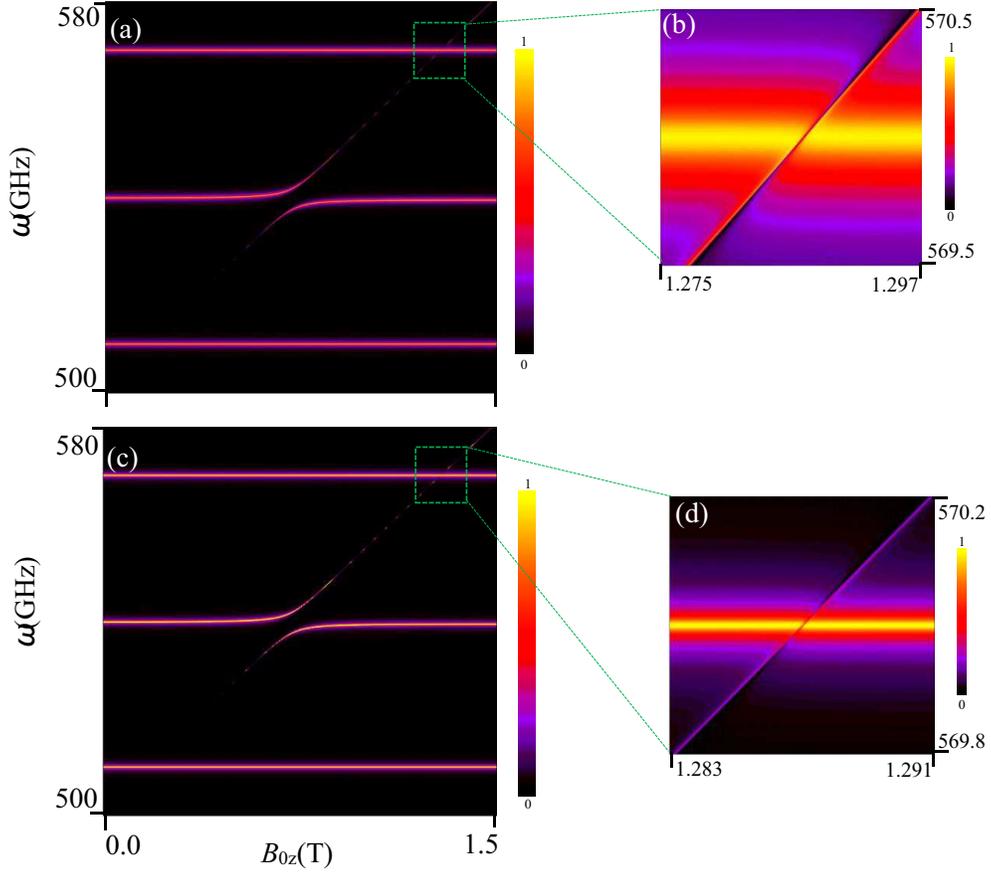}
\caption{ $|S_{21}|$ transmission spectra of a 0.1$\mu m$-thick DAI obtained from the numerical method. The regions in green box of left panels are zoomed-in and shown in the right panels. Top (\textbf{a} and \textbf{b}) and bottom (\textbf{c} and \textbf{d}) panels are for $R$=0.99 and $R$=0.999.  } \label{fig3}
\end{figure*}

\begin{figure*}[!ht]
\centering
\includegraphics[angle=0,scale=0.8]{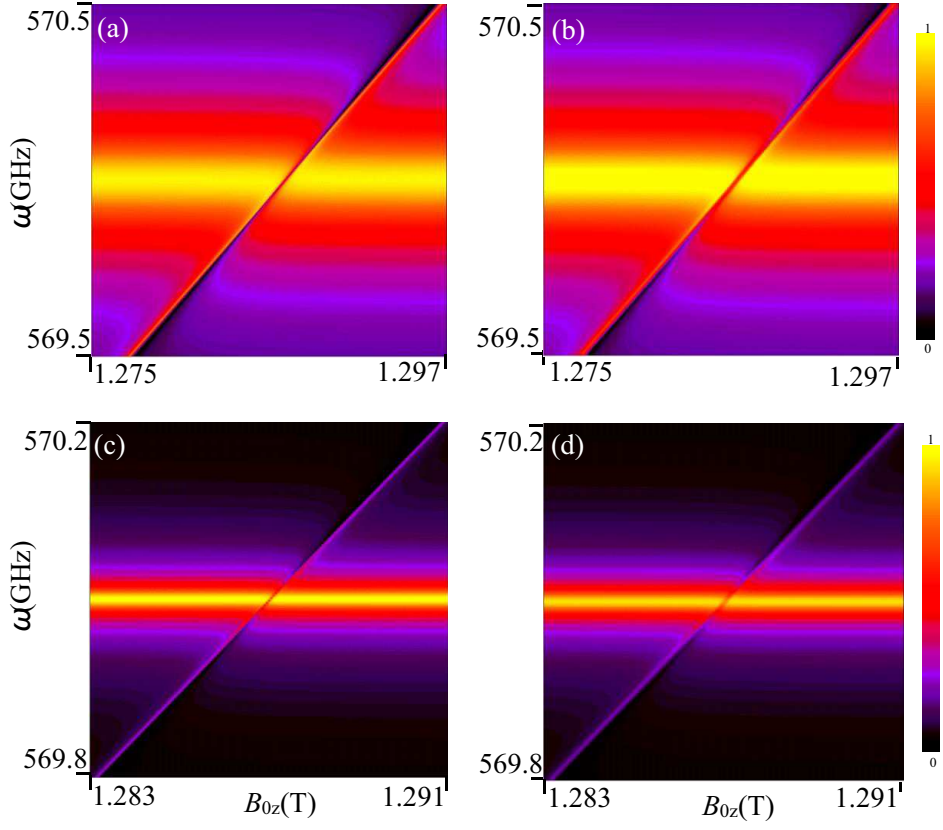}
\caption{ $|S_{21}|$ transmission spectra of a 0.1$\mu m$-thick DAI for odd mode obtained from (left) the numerical method and (right) the formula Eq. (\ref{S21o2m}). Top (\textbf{a} and \textbf{b}) and bottom (\textbf{c} and \textbf{d}) panels are for $R$=0.99 and $R$=0.999. } \label{fig4}
\end{figure*}

\begin{figure*}[t]
\centering
\includegraphics[angle=0,scale=0.95]{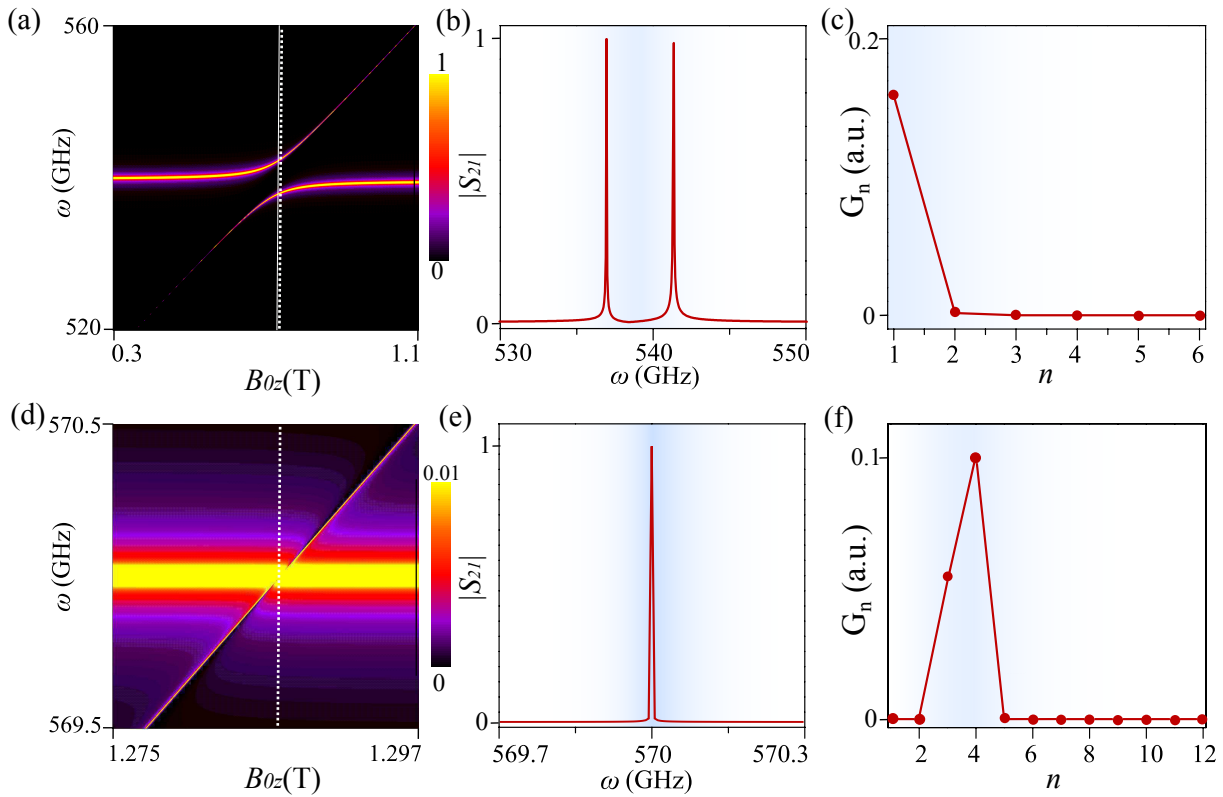}
\caption{  (\textbf{a}) $|S_{21}|$ transmission spectra of the even mode as a function of the static magnetic field and the photon frequency, which is calculated based on Eq. (\ref{S21em}) for the DAI film thickness $l_s=0.1\mu m$. The strong coupling manifests the LR. (\textbf{b}) Spectrum at resonant magnetic field represented by the white dashed line in (\textbf{a}). (\textbf{c}) Absolute value of the term $G_n$ in the expansion of $G= \sum\limits_{n=0}^{+\infty} G_{n}$. As for the even mode, one can see that the linear term ($n$=1) dominates. (\textbf{d-f}) The same as (\textbf{a}-\textbf{c}) but for the odd mode and calculated using Eq. (\ref{S21o1m}). In order to avoid divergence, a imaginary part is added into the frequency $\omega$. As for the odd mode, the first- and  second-order terms are vanishing, indicating the high-order nature of the coupling. } \label{fig5}
\end{figure*}

\begin{figure*}[t]
\centering
\includegraphics[angle=0,scale=0.8]{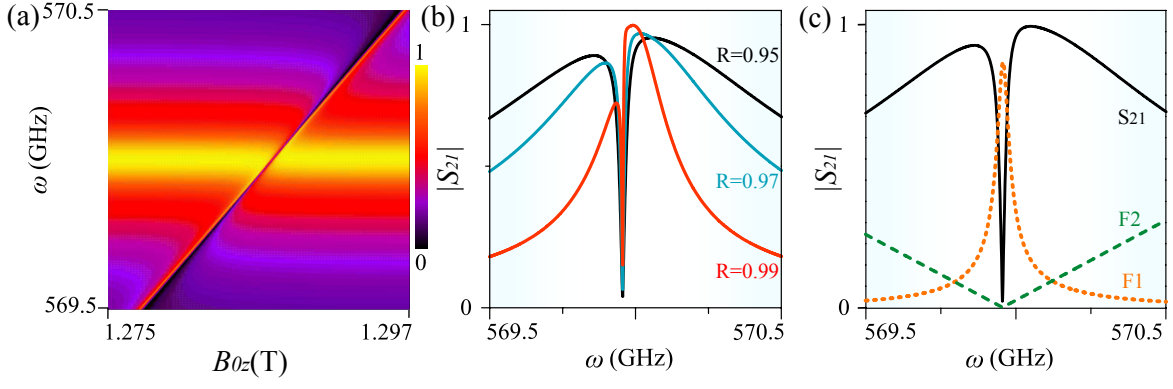}
\caption{  (\textbf{a}) $|S_{21}|$ transmission spectra of the odd mode calculated with the numerical method (without using Taylor expansion) for a DAI film thickness $l_s=0.1\mu m$ and the reflectivity $R=0.99$ of the cavity. The reflectivity $R$ of electromagnetic wave at two ports determines the decay rate of intracavity field. (\textbf{b}) $|S_{21}|$ spectra for the reflectivity $R$=0.95 (black), 0.97 (blue), 0.99 (red) at the resonant magnetic field.  As the reflectivity $R$ increases, the dissipation is gradually suppressed, and the gap width is enhanced little by little. But we notice that the position of the gap remains unchanged. (\textbf{c}) The schematic mechanism of gaped LA. The approximated formula of  $S_{21}$ spectra is written as $(\omega-m')/[(\omega-\omega'_c)(\omega-\omega'_1)]$ (S$_{21}$,black), which can be considered as the combination of two functions $1/[(\omega-\omega'_c)(\omega-\omega'_1)]$ (F1, brown) and $(\omega-m')$ (F2, green). Here, `$m'$, $\omega'_c$ and $\omega'_1$' are renormalized by the dissipation.  } \label{fig6}
\end{figure*}

\begin{figure*}[t]
\centering
\includegraphics[angle=0,scale=0.8]{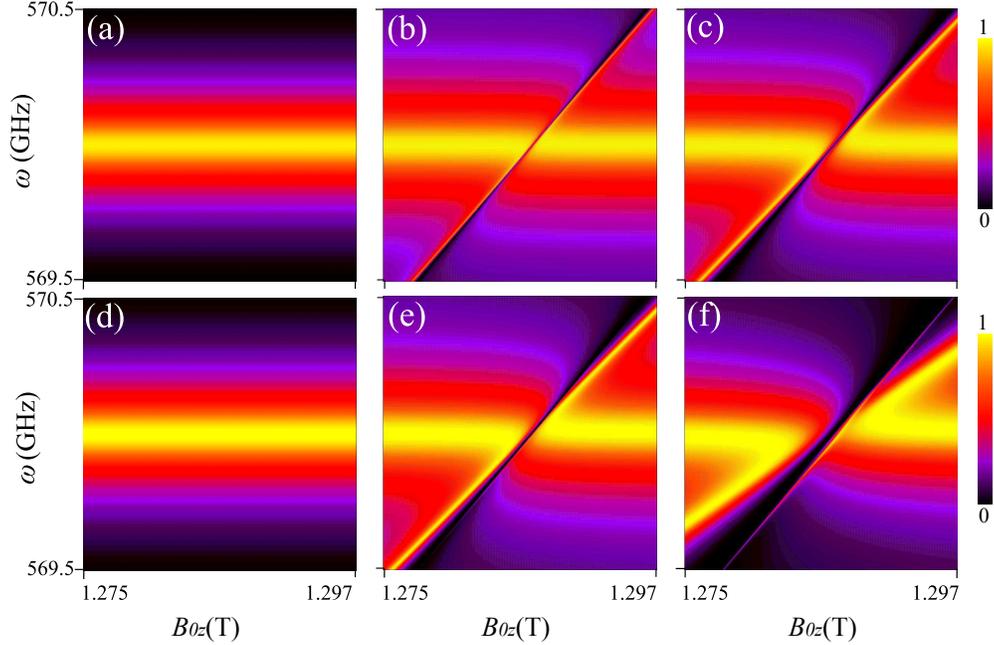}
\caption{ $|S_{21}|$ transmission spectra for a DAI film thickness $l_s$=0.1 $\mu m$ with different coupling strength $b$= 0(\textbf{a}), 1(\textbf{b}) and 2(\textbf{c}) meV. Controllable coupling strength $b$ can be obtained through applying the external magnetic field in topologically nontrivial DAI. We can see that an enhanced gap in the LA is induced by strong coupling. (\textbf{d-f}) The same as (\textbf{a}-\textbf{c}), but for a thicker DAI film $l_s$=0.5 $\mu m$. Note that high-order modes arise in the spectrum for which the high-order interaction dominates. For all data here, the reflectivity $R$ is fixed to be 0.99. } \label{fig7}
\end{figure*}

\begin{figure*}[!ht]
\centering
\includegraphics[angle=0,scale=0.7]{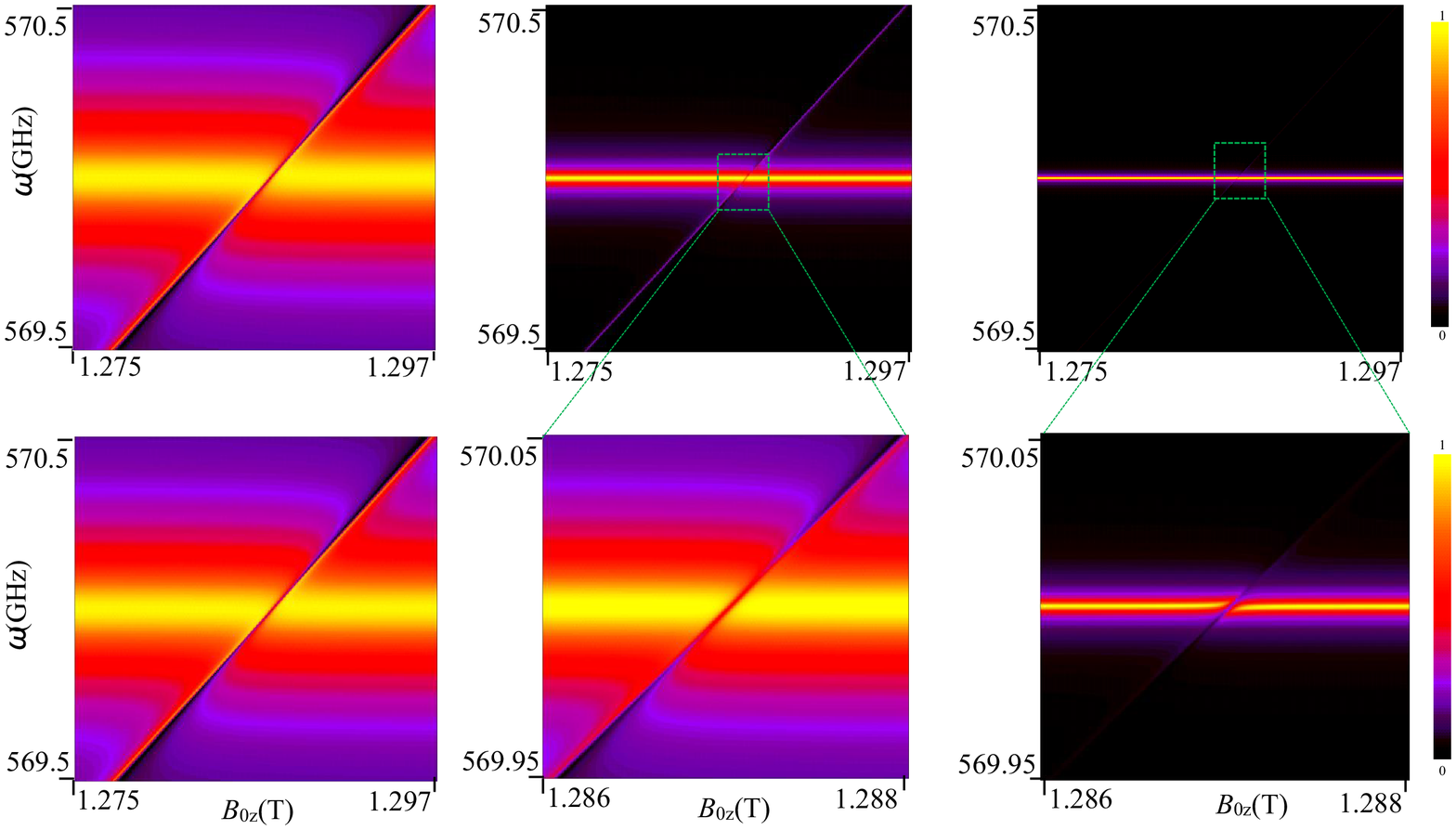}
\caption{ $|S_{21}|$ transmission spectra of a 0.1$\mu m$-thick DAI for odd mode calculated with numerical method for $R$=0.99 (left), $R$=0.999 (middle) and $R$=0.9999 (right). The regions in green box of top panels are zoomed-in and shown in the bottom panels. Notice that two figures of $R$=0.99 (left top and left bottom) are the same for the sake of completeness.} \label{fig8}
\end{figure*}


\section{I. Introduction}
Axion was first postulated as an elementary particle to solve the charge-parity puzzle in the strong interaction between quarks in the particle physics~\cite{peccei1977}. But its existence in nature is another puzzle. Interestingly, axion emerges  as a quasiparticle in three-dimensional (3D) topological insulators through the axion action $\mathcal{S}_{\text{topo}}=(\theta/2\pi)(e^2/hc_0)\int d^3xdt \mathbf{E}\cdot{\mathbf{B}}$ from the topological field theory~\cite{qi2011,qi2008}, in which $\mathbf{E}$ and $\mathbf{B}$ are the electric field and magnetic induction inside the insulators, $e$ is the charge of an electron, $h$ is Plank's constant, $c_0$ is the speed of light in the vacuum, $\theta$ (modulo $2\pi$) is the dimensionless pseudoscalar parameter as the axion field. Once both time reversal symmetry and inversion symmetry are broken, the axion field $\theta$ becomes dynamical due to the spin-wave excitation, denoted as a dynamical axion field (DAF) $\theta(\bm{r},t)$~\cite{li2010,wang2011prl}.   Many exotic electromagnetic phenomena were predicted for DAF systems, such as, axion polariton~\cite{li2010,marsh2018prl}, the chiral magnetic effect~\cite{sekine2016,sumiyoshi2016}, and so on~\cite{ooguri2012, taguchi2018,imaeda2019jpsj, gooth2019}.

Recently, several topological dynamical axion insulators (DAIs) were proposed to a host large DAF characterized by a nonzero spin Chern number~\cite{wang2020dynamical} and strongly couple with the light,  for example, van der Waals layered material Mn$_2$Bi$_2$Te$_5$~\cite{zhang2020cpl} and MnBi$_2$Te$_4$/Bi$_2$Te$_3$ superlattice~\cite{wang2020dynamical}. Especially, some of us predicted tunable DAF in MnBi$_2$Te$_4$ films~\cite{zhu2020} which have been synthesized in experiments~\cite{gong2019cpl,otrokov2019nature,deng2020,liu2020robust,chen2019intrinsic,klimovskikh2020npj}. The proposed DAIs provide a guarantee for the in-depth study of axion electrodynamics~\cite{nenno2020,sekine2020axion}. The axion polariton is one of the most interesting electromagnetic phenomena, where the axion quasiparticle in DAIs couples with free photons~\cite{li2010}. However, the tunability of the free-photon setup turns out to be limited and no experimental evidence of axion polariton is reported. Compared to free photons, cavity photons provide a better choice due to large quality factor and enhanced light-matter interaction~\cite{cohentannoudji2004book,haroche2013rmp,savage1989prl}. Notably, the rapid progress in cavity fabrication has already made the THz cavity possible~\cite{scalari2012science}, which covers the typical energy scale ($\sim 1$ meV) of axion quasiparticles in DAIs~\cite{li2010}. This enables to study the axion polariton in a cavity, dubbed as cavity axion polariton. One expects that the cavity axion polariton will present unique properties compared to the axion polariton with free photons~\cite{li2010}.

In this paper, we studied the electrodynamics of cavity axion polariton by embedding a DAI into a THz cavity. Intriguingly, in addition to the usual level repulsion (LR)~\cite{aspelmeyer2014rmp,agarwal1984prl,tabuchi2015science}, schematically shown in Fig.~\ref{fig1}(a), a characteristic level attraction (LA), schematically shown in Fig.~\ref{fig1}(c), was found. Based on numerical and analytical calculation, we found that such an LA originates from the high-order axion-photon interaction for which we call it high-order LA. This new mechanism, which is distinct from linear-order coupling of conventional LA~\cite{metelmann2015prx,bernier2018pra,harder2018prl,yu2019prl,tserkovnyak2005rmp,bhoi2019prb,grigoryan2018prb,boventer2020prr} schematically shown in Fig. \ref{fig1}(b), has so far not been revealed.

The remainder of this paper is organized as follows. In Sec. II, we provide a comprehensive description of theoretical and methodological details used in this work. For the convenience of readers, the detailed derivation of some formulae can be found in Appendix A$\sim$C. Section III is devoted to the presentation and interpretation of our numerical and analytical results for the high-order LA.

\section{II. Theory and method}
\subsection{IIA. Axion electrodynamics}
The axion-photon coupled system can be described by the effective action~\cite{li2010}
\begin{equation}
\begin{split}
\mathcal{S}_{\mathrm{tot}}=& \mathcal{S}_{\text { Maxwell }}+\mathcal{S}_{\text { topo }}+\mathcal{S}_{\text { axion }} \\
&= \frac{1}{8 \pi} \int d^3xdt \left(\mathbf{D} \cdot \mathbf{E}-  \mathbf{B} \cdot \mathbf{H} \right) \\
&+\frac{\alpha}{4 \pi^{2}}  \left(\theta_{0}+\delta \theta\right) \int d^3xdt \mathbf{E} \cdot \mathbf{B}\\
&+g^{2} J \int d^3xdt \left[\left(\partial_{t} \delta \theta\right)^{2}-\left(v_{i} \partial_{i} \delta \theta\right)^{2}-m^{2} \delta \theta^{2}\right], \label{Saction}
\end{split}
\end{equation}
where $\mathbf{H}$ and $\mathbf{D}$ are the vectors of magnetic field and electric displacement. The constitutive relations are $\mathbf{B}=\mu_r \mathbf{H}$ and $\mathbf{D}=\epsilon \mathbf{E}$ with $\epsilon$ and $\mu_r$ the dielectric constant and magnetic permeability. $\alpha=e^{2}/\hbar c_0$ is the fine-structure constant, the axion field $\theta = \theta_{0}+\delta \theta$ with $\theta_0$ the static part of the axion field. $\delta\theta$ represents the DAF originating from the longitudinal spin-wave excitation with material-dependent stiffness $J$, velocity $v_i$, mass $m$, and coefficient $g$~\cite{li2010}. As given by the $\mathcal{S}_{\text { topo }}$ term, strong coupling between axion field ($\delta \theta$) and electromagnetic field ($\mathbf{E}$) can be realized by applying a static magnetic field $\mathbf{B}_0$ parallel to the electric field $\mathbf{E}$.

In order to obtain the dynamical equations for the coupled axion-photon system, we employ the Euler-Lagrangian equation and obtain the modified Maxwell's equations Eqs. (A1)$\sim$(A4) and the equation of motion for the axion field Eq. (A5). After some algebra and calculus, we further obtain
\begin{eqnarray}
\frac{\partial^2 \mathbf{E}}{\partial t^2} - c^2 \bigtriangledown^2 \mathbf{E} + \frac{\alpha}{\pi \epsilon} \frac{\partial^2 \delta \theta}{\partial t^2} \mathbf{B}_0 = 0, \label{maxwell6m}\\
\frac{\partial^2 \delta \theta}{\partial t^2} + m^2 \delta \theta + \Gamma \frac{\partial \delta \theta}{\partial t} - \frac{\alpha }{8\pi^2 g^2 J} \mathbf{B}_0 \cdot \mathbf{E} = 0, \label{DAF2m}
\end{eqnarray}
where the speed of light in the DAI is $c=c_0/\sqrt{\epsilon\mu_r}$. $\Gamma$ is the intrinsic damping of axion mode. Equation (\ref{DAF2m}) also implies that the mass $m$ of DAF provides the resonance frequency of axion mode. In this work, the axion field is uniform inside the DAI and the axion mode is of a very long wavelength. Therefore, the dispersion of axion mode is negligible and is henceforth not taken into account. Moreover, the magnetic induction $\mathbf{B}$ in Eq. (\ref{Saction}) consists of both magnetic component of electromagnetic wave $\mathbf{B}_{ac}$ and static magnetic field $\mathbf{B}_0$. However, the former is in general four orders of magnitude smaller than the latter and thus only $\mathbf{B}_0$ is present in the above two equations.

\subsection{IIB. Axion mode and parallel pumping}
In contrast to the dark axion in the universe, the axion mode studied in this work is a quasiparticle existing in an antiferromagnetic DAI. As predicted in Ref.~\cite{li2010}, the axion quasiparticle arises from the longitudinal spin wave excitations. That is to say, $\delta \theta \propto \delta M_z^-$ with $M_z^-=(M_z^A - M_z^B)/2$ the $z$ component difference of two sublattice magnetizations $\mathbf{M}^{A,B}$ of an antiferromagnet. The saturation magnetizations of two sublattices are aligned along the $z$ direction. In general, there are a number of thermal spin wave inside the DAI but its tunability is uneasy. To better excite and manipulate the spin wave, we consider the light-driven spin wave which can be controlled by tuning the frequency and the power of light. There are two ways of exciting spin wave, i.e. perpendicular pumping~\cite{huebl2013prl,tabuchi2014prl,goryachev2014prapplied,zhang2014prl,bai2015prl,wang2018prl} and parallel pumping~\cite{morgenthaler1963prl}. In the former case, the magnetic component $\mathbf{B}_{ac}$ of electromagnetic wave is perpendicular to the static magnetic field $\mathbf{B}_0$. Hence, only transverse magnetization $M_{x,y}^{A,B}$ is excited. Both ferromagnetic resonance (FMR) and antiferromagnetic resonance (AFMR) belong to perpendicular pumping~\cite{huebl2013prl,tabuchi2014prl,goryachev2014prapplied,zhang2014prl,bai2015prl,wang2018prl}. The resonance frequency usually takes the form of $\omega^{\bot} = \gamma B_{0z}$ for FMR and $\omega^{\bot} = \omega_0 \pm \gamma B_{0z}$ for AFMR. $\gamma$ is the the gyromagnetic ratio, $B_{0z}$ is the $z$ component of $\mathbf{B}_0$ and $\omega_0$ is the zero-field frequency of AFMR. Due to two sublattice magnetizations, two magnon modes and magnon frequencies occur in AFMR.

As for parallel pumping, $\mathbf{B}_{ac}$ is parallel to static magnetic field and two sublattice magnetizations. The time-varying nature of $\mathbf{B}_{ac}$ will induce the periodically-precessing magnetization and result in $\delta M_z^-$ and $\delta \theta$. The frequency of $\delta M_z^-$ variation in parallel pumping is given by $\omega^{\parallel} = 2\omega^{\bot}$~\cite{morgenthaler1963prl}. From a quantum point of view, the parallel pumping is a process that one photon splits into two magnons with the same frequency. Therefore, the resonance frequency of axion mode in Eq. (\ref{DAF2m}) is written as $m = 2\omega_0 \pm 2\gamma B_{0z}$. In this work, we consider the case of large magnetic field and thus two frequencies of axion mode are separated. So, two axion modes can be regarded as independent. In the following, only the axion mode with the frequency $m = 2\omega_0 + 2\gamma B_{0z}$ is considered in this work.

\subsection{IIC. Axion-photon coupling strength}

As described in the previous section, the dynamic magnetic field $\mathbf{B}_{ac}$ is aligned along the $z$ axis in order to give rise to $\delta M_z$ and $\delta \theta$. So, the dynamic electric field $\mathbf{E}$ is perpendicular to the $z$ axis. We assume that $\mathbf{E}$ is along the $x$ axis. One can see from Eq. (\ref{DAF2m}) that the term of $\mathbf{B}_0 \cdot \mathbf{E}$ should be nonzero for the axion-photon coupling. This requires that $\mathbf{B}_0$ should have an $x$ component. On the other hand, the frequency of axion mode is related to the $z$ component of $\mathbf{B}_0$. For these two reasons, we apply a static magnetic field $\mathbf{B}_0$ that is oriented with an angle $\varphi$ with respect to the $z$ axis in the $x$-$z$ plane. In such case, Eqs. (\ref{maxwell6m}) and (\ref{DAF2m}) can be rewritten in scalar forms as
\begin{eqnarray}
\frac{\partial^2 E}{\partial t^2} - c^2 \bigtriangledown^2 E + \frac{\alpha B_{0x}}{\pi \epsilon} \frac{\partial^2 \delta \theta}{\partial t^2}  = 0, \label{maxwell7m} \\
\frac{\partial^2 \delta \theta}{\partial t^2} + m^2 \delta \theta + \Gamma \frac{\partial \delta \theta}{\partial t} - \frac{\alpha B_{0x}}{8\pi^2 g^2 J} E = 0. \label{DAF3m}
\end{eqnarray}

By assuming the time- and position-dependence of $e^{-j\omega t + j \mathbf{k} \cdot \mathbf{r}}$ for $E$ and $\delta \theta$, one can simultaneously solve Eqs. (\ref{maxwell7m}) and (\ref{DAF3m}) and obtain the eigenfrequency
\begin{widetext}
\begin{equation}
\omega_{\pm}^2 = \frac{ (c^2k^2+m^2+b^2-j\Gamma\omega) \pm \sqrt{ (c^2k^2+m^2+b^2-i\Gamma\omega)^2  - 4c^2k^2(m^2-j\Gamma\omega)  } }{2} \label{eigenW}
\end{equation}
\end{widetext}
where $b=\sqrt{\alpha^{2} B_{0x}^{2} /8 \pi^{3} \epsilon g^{2} J}$. Equation (\ref{eigenW}) shows that the axion-photon coupling results in two polariton modes with frequency $\omega_{\pm}$. At $k$=0 and $\Gamma$=0, $\omega_+=\sqrt{m^2+b^2}$ and $\omega_-=0$. As $k \rightarrow +\infty$, $\omega_+ \rightarrow ck$ and $\omega_- \rightarrow m$. Thus, there is an anticrossing gap between $m$ and $\sqrt{m^2+b^2}$ in which the wave can not propagate. This is in agreement with phonon, exciton and magnon polaritons~\cite{cohentannoudji2004book,haroche2013rmp,savage1989prl}.

Based on Eq. (\ref{eigenW}), the size of anticrossing gap in the spectrum is related to the parameter $b$. The larger the parameter $b$ is, the larger the gap size is and the stronger the axion-photon coupling is. So, an important task is to determine the value of parameter $b$. To do so, we consider the recently proposed MnBi$_2$Te$_4$ DAIs by some of the authors~\cite{wang2020dynamical,zhang2020cpl,zhu2020}, which are predicted to host a large DAF and may be potential candidates for observing the axion polariton.

The low-energy physics of the MnBi$_2$Te$_4$-based DAIs can be described by the following four-band Dirac Hamiltonian~\cite{wang2020dynamical, zhang2020cpl}
\begin{equation}
\label{dkgamma}
\begin{split}
\mathcal{H}_{0}(\mathbf{k}^B)=&E_{0}(\mathbf{k}^B)\mathbb{I}_{4\times 4}+\sum^{5}_{i=1}d_{i}(\mathbf{k}^B)\Gamma^{i},\\
d_{1,2,...,5}(\mathbf{k}^B)=&\Big(A_{2}k_y^B,-A_{2}k_x^B,A_{1}k_z^B,M^B(\mathbf{k}^B), m_5 \Big),
\end{split}
\end{equation}
where $\mathbf{k}^B$ is the wave vector in the Brillouin zone. $E_{0}(\mathbf{k})=C+D_{1}(k_z^B)^{2}+D_{2}((k_x^B)^{2}+(k_y^B)^{2})$, $M^B(\mathbf{k})=M+B_{1}(k_z^B)^{2}+B_{2}((k_x^B)^{2}+(k_y^B)^{2})$, and the five Dirac matrices are represented as $\Gamma^{1,2,...,5}=(\tau_{x}\otimes\sigma_{x},\tau_{x}\otimes\sigma_{y},\tau_{y}\otimes \sigma_{0}, \tau_{z}\otimes \sigma_{0},\tau_{x}\otimes\sigma_{z})$, which satisfy the Clifford algebra $\{\Gamma^{i},\Gamma^{j}\}=2\delta_{i,j}$.  The parameters $B_{i=1,2}$ are assumed to be positive, and $M$ is the mass term related to spin-orbit coupling with $M<0$ ($M>0$) corresponding to the topological (trivial) antiferromagnetic insulator with(without) band inversion. The $m_5\Gamma_5$ term results from the antiferromagnetic order, which breaks both time-reversal and inversion symmetries and is essential for the emergence of dynamical axion states. For convenience, in the numerical calculations, we choose the same values of the parameters $A_{i}, B_{i}, C, D_{i}$ and $m_5$ as those in Ref. \cite{wang2020dynamical} for MnBi$_2$Te$_4$-Bi$_2$Te$_3$ heterostructure with $C=0.0232$ eV, $D_{1}=1.77$ $\mathrm{eV\ \AA^{2}}$, $D_{2}=10.82$ $\mathrm{eV\ \AA^{2}}$, $A_{1}=0.30$ $\mathrm{eV\ \AA}$, $A_{2}=1.76$ $\mathrm{eV\ \AA}$, $B_{1}=2.55$ $\mathrm{eV\ \AA^{2}}$, $B_{2}=14.20$ $\mathrm{eV\ \AA^{2}}$, $m_{5}=3.8$ meV. It is worth mentioning that the mass parameter $M$ can be tuned by changing the spin-orbit coupling strength through chemical doping, and is thus treated as a variable in our calculations.

The value of $\theta$ in this model can be simply calculated via the formula~\cite{li2010}
\begin{equation}
\theta=\frac{1}{4\pi}\int d^{3}k\frac{2|d|+d_{4}}{(|d|+d_{4})^{2}|d|^{3}}\epsilon^{ijkl}d_{i}\partial_{x}d_{j}\partial_{y}d_{k}\partial_{z}d_{l}, \label{DAF4}
\end{equation}
where $i,j,k,l$ take values from 1, 2, 3, 5 and $|d|=\sqrt{\sum_{n=1}^{5}d_{n}^{2}}$ with the lattice-regularized components of $d_{1,2,...,5}$ in the continuum model. Figure~\ref{fig2}(a) shows $\theta$ as a function of $M$, which approaches $\pi$ (0) for topologically nontrivial (trivial) phase with $M<0$ ($M>0$).  In the presence of longitudinal magnetization fluctuations, the axion field becomes dynamical, and to the linear order, $\delta\theta(\mathbf{x},t)=\delta m_{5}/g$, where $\delta m_{5}$ is proportional to the amplitude fluctuation of antiferromagnetic order along the $z$-direction $\delta M_z$, and the coefficient  $1/g$ is determined from Eq. (\ref{DAF4}). In Fig.~\ref{fig2}(b), we plot $1/g$ as a function of $M$, and found that $1/g$ is significantly enhanced near the topological transition point $M \rightarrow 0^-$.

In addition, the stiffness parameter $J$ is obtained from the above effective model as \cite{li2010}
\begin{equation}
J=\int \frac{\mathrm{d}^{3} k}{(2 \pi)^{3}} \frac{d_{i}(\mathbf{k}) d^{i}(\mathbf{k})}{16|d|^{5}}\sim\int \frac{\mathrm{d}^{3} k}{(2 \pi)^{3}} \frac{1}{16|d|^{3}}=\frac{1}{\Omega}\sum_{\mathbf{k}}\frac{1}{16|d|^{3}},
\end{equation}
where $\quad (i=1,2,3,4)$ and $\Omega$ is the size of the unit cell. Figure~\ref{fig2}(c) shows the value of $J$ as a function of $M$. Taking the values of $J$ and $1/g$ together, we plot the band-structure-dependent coefficient $1/(g\sqrt{J})$ of $b$ as a function of $M$ in Fig.~\ref{fig2}(d), where the largest magnitude is achieved near the topological transition.

Finally, after choosing typical values of the permittivity $\epsilon\sim20$ and the static magnetic field $B_{0z} \sim 0.2$ T, the realistic axion-photon coupling strength $b$ as a function of $M$ can be evaluated. As shown by Fig.~\ref{fig1}(e), it is distinguished between topologically nontrivial ($M<0$) and trivial ($M>0$) DAIs. The magnitude of the coupling strength $b$ in the nontrivial regime is significantly larger than that in the trivial regime. The strongest axion-photon coupling can be obtained near the topological transition ($M\sim0$). Therefore, an antiferromagnetic topological insulator with a large topological DAF is an ideal candidate for studying the strong axion-photon coupling in a cavity.

\subsection{IID. Transmission spectrum}

Transmission spectrum is an important tool of studying the light-matter interaction in both experimental and theoretical works~\cite{huebl2013prl,tabuchi2014prl,goryachev2014prapplied,zhang2014prl,bai2015prl,wang2018prl}. To discuss the axion-photon coupling, we calculate the transmission spectrum of a DAI-embedded cavity shown in Fig.~\ref{fig1}(d). The axis of cavity and the propagation direction of electromagnetic waves are along the $y$ axis. Due to the axion-photon coupling, the photon transmission $S_{21}$ carries the information of photon, axion and the coupling between them. The cavity is a one-dimensional hollow cylinder with input and output ports at the left and right ends. The input port is driven by external sources with an incident wave amplitude and the output port is detected with a transmitted wave amplitude. The transmitted-to-incident ratio is defined as transmission $S_{21}$. Inside a cavity, the wave is reflected many times. Thus the left-going waves and the right-going waves are superposed to form a standing wave. The cavity modes consist of many discrete resonance modes, which can be either even mode or odd mode as shown in Fig.~\ref{fig1}(d)~\cite{pozar2011book}.  An antiferromagnetic DAI with the thickness of $l_s$ is placed in the middle of the cavity. As for the even mode, the amplitude of electric field $\mathbf{E}$ at the DAI is large and thus the axion-photon coupling is strong. In contrast, the odd mode is weakly coupled with the DAI.


To calculate the transmission $S_{21}$, we first calculate the wave vector $k$ of propagation state inside the DAI. From Eq. (\ref{eigenW}), one can give rise to the relation $k(\omega)$
\begin{equation}
c^2k^2 = \omega^2 \mu, \label{k1}
\end{equation}
with the parameter
\begin{align}
\mu = \frac{\omega^2 +i\Gamma \omega -(b^2+m^2)}{\omega^2 +i\Gamma \omega -m^2}. \label{mumu}
\end{align}
Here, the parameter $\mu$ does not refer to the permeability.

With the wave vector $k$, one can calculate the impedance $Z = \frac{E}{H} = -\frac{\omega }{k}$ with the help of Eq. (\ref{maxwell3}). Inside the cavity, there exist three regions, i.e. the left air region, the DAI and the right air region. Therefore, we have three transfer matrices which connect the electromagnetic field at the left surface and right surface of each region. As given in Appendix B, the transfer matrix in the DAI region is written as
\begin{equation}
T_{DAI} =
\left( \begin{array}{cc}
           \cos(kl_s) \ \ \ \ \ \ -jZ\sin(kl_s) \\
           -\frac{j}{Z}\sin(kl_s) \ \ \ \ \ \ \cos(kl_s) \\
\end{array} \right) \label{transferDAIm}
\end{equation}
where $Z=-\frac{c_0}{\sqrt{\epsilon\mu}}$ and $k=\frac{\omega}{c_0}\sqrt{\epsilon\mu}$.

The transfer matrix of the air region $T_{air}$ is obtained by replacing the quantities $k$, $Z$ and $l_s$ with those of the air region, i.e. $k_0=\frac{\omega}{c_0}$, $Z_0=-c_0$ and $l=\frac{L-l_s}{2}$. $l$ defines the length of two air regions and $L$ defines the total length of the cavity. The total transfer matrix of a cavity is determined by multiplying the transfer matrix of each region, i.e. $T=T_{air}T_{DAI}T_{air}$.

With the transfer matrix $T$, we determine the relation between the amplitudes of incident wave at input port and those of transmitted wave at output port~\cite{pozar2011book,yao2015prb}
\begin{equation}
\left( \begin{array}{c}
           E_A^- \\
           E_B^- \\
\end{array} \right) =
S \left( \begin{array}{c}
           E_A^+ \\
           E_B^+ \\
\end{array} \right) =
\left( \begin{array}{cc}
           S_{AA} \ \ \ \ S_{AB} \\
           S_{BA} \ \ \ \ S_{BB} \\
\end{array} \right)
\left( \begin{array}{c}
           E_A^+ \\
           E_B^+ \\
\end{array} \right), \label{SC1m}
\end{equation}
where $-(+)$ represents the direction from the cavity (outside) to the outside(cavity). The detailed form of scattering matrix $S$ is given in Appendix B.

In order to create standing waves inside a cavity, strong reflections at the input and output ports are necessary. We define the reflectivity of intracavity photon at two ports by the parameter $R$ and we choose $R\approx$1 for both ports~\cite{pozar2011book,yao2015prb}.

The transmission is written as
\begin{align}
S_{21} = \frac{ (1-R^2) S_{BA}}{(1-RS_{AA})(1-RS_{BB})- R^2S_{AB}S_{BA}}. \label{S21m}
\end{align}

Substituting the elements of scattering matrix Eq. (\ref{SC1m}) into Eq. (\ref{S21m}), we obtain
\begin{align}
S_{21} = \frac{F}{G}, \label{S21expan}
\end{align}
where the numerator $F$ and the denominator $G$ are given in Appendix B.


In the numerical calculations, we adopt the values of typical parameters which are within the achievable range of experimental measurement. We take the dielectric constant $\epsilon$=20, the permeability $\mu_r$=1, the length of cavity $L$=5 mm, the axion frequency at zero magnetic field $\omega_0$=1 meV, and the intrinsic damping of axion $\Gamma$=0.01 meV. The parameters used in the tight-binding calculations are given in Sec. IIC. The other parameters, e.g. the thickness $l_s$ of DAI, the coupling strength $b$, the static magnetic field $B_{0z}$, vary and are shown in the figures.

\subsection{IIE. Analytical expressions}

In order to better understand the numerical results, we derive the analytical expressions of transmission $S_{21}$ and present simple formulae with transparent physics. The analytic expression of $S_{21}$ is obtained by performing the Taylor expansion around cavity resonance frequency $\omega_c$ and in the limit of long-wavelength with $kl_s \ll 1$.

First, the sinusoidal functions $\sin(kl_s)$ and $\cos(kl_s)$ in transfer matrix and scattering matrix are expanded as
\begin{eqnarray}
\sin (k l_s) &=&  \sum\limits_{n=0}^{+\infty} \frac{(-1)^n}{(2n+1)!} (k l_s)^{2n+1}, \label{sin1} \\
\cos (k l_s) &=&  \sum\limits_{n=0}^{+\infty} \frac{(-1)^n}{(2n)!} (k l_s)^{2n}. \label{cos1}
\end{eqnarray}

Second, the exponential function $e^{-2jk_0l}$ in the scattering matrix is expanded at cavity resonance frequency $\omega_{CR}=\frac{c_0q\pi}{2l+l_s}$, i.e.
\begin{align}
e^{-2jk_0l} &= e^{-j \frac{2l}{c_0} \omega} = e^{j \frac{l_s}{c_0} \omega} e^{-j \frac{2l+l_s}{c_0} \omega} \notag \\
            &\approx e^{ -j \frac{2l+l_s}{c_0} \omega_{CR} } e^{j \frac{l_s}{c_0} \omega} \left[ 1 + \Delta \omega_C + \cdots \right] \notag \\
            &= e^{ -j q \pi } \sum\limits_{n_1}^{+\infty} \frac{(j \frac{l_s}{c_0} \omega)^{n_1}}{n_1!}  \sum\limits_{n_2}^{+\infty} \frac{(\Delta \omega_C)^{n_2}}{n_2!} \label{exp1}
\end{align}
where $\Delta \omega_C = (-j \frac{2l+l_s}{c_0}) (\omega-\omega_{CR})$.

Substituting Eqs. (\ref{sin1})$\sim$(\ref{exp1}) into Eqs. (\ref{S21m}), one can obtain the analytical expressions of the numerator $F$ and the denominator $G$. The form of numerator $F$ is simple and its analytical expression can be easily derived. However, the expression of denominator $G$ is lengthy and cumbersome and will be discussed in detail.

\subsubsection{Even mode}
As for the even mode, i.e. the integer $q$ in Eq. (\ref{exp1}) is even, the denominator is written as $G^e = \sum\limits_{n_1=0}^{+\infty} G_{n_1}^e$ with $G_{n_1}^e = \sum\limits_{n_2=0}^{+\infty} G_{n_1,n_2}^e (\Delta \omega_C)^{n_2} (\frac{l_s}{c_0} \omega)^{n_1}$. The detailed expansion of $G$ is given in Appendix C. For the sake of analysis, we divide the terms in $G$ into two groups. The first group, which is called the type-I interaction, consists of the terms without $\Delta \omega_C$, i.e. those terms with $n_2$=0. The second group is called type-II interaction and contains all remaining terms.

As $l_s$ is small, the denominator $G^e$ is dominated by the zeroth-order term and the linear-order term of the type-I interaction, i.e.
\begin{align}
G^e &= -8 \Delta \omega_C + 8j ( \frac{l_s}{c_0} \omega ) (\epsilon \mu - 1) \notag \\
&= 16j\frac{l}{c_0} \left[ (\omega-\omega_C) - \frac{\epsilon b^2 l_s}{4l(\omega-m)}  \right]. \label{even-denm}
\end{align}

With the numerator $F$, the transmission $S_{21}$ of even mode for small $l_s$ is written as
\begin{align}
S_{21}^{e} = C^{e} \frac{(\omega-m)}{(\omega-\omega_c)(\omega-m)-g^2_e}, \label{S21em}
\end{align}
where $C^{e}$ denotes a constant and $g^2_e=\epsilon l_sb^2/{4l}$.


\subsubsection{Odd mode} As for the odd mode, i.e. the integer $q$ in Eq. (\ref{exp1}) is odd, the denominator is written as $G^o = G_x + \sum\limits_{n_1=0}^{+\infty} G_{n_1}^o$ with $G_{n_1}^o = \sum\limits_{n_2=0}^{+\infty} G_{n_1,n_2}^o (\Delta \omega_C)^{n_2} (\frac{l_s}{c_0} \omega)^{n_1}$.

By comparing the expansion of even mode and odd mode given in Appendix C, we found that the 1$^{st}$- and 2$^{nd}$-order terms of type-I interaction are absent for the odd mode. Therefore, the high-order type-II terms dominate, resulting in a high-order LA. As $l_s$ is small, the dominant terms are those with $G_{1,1}^o$ and $G_{2,2}^o$, i.e. Eq. (C7) and (C8), and the term $G_x$, i.e. Eq. (C11), which are validated by numerical calculations.

In order to distinguish the role played by each term, we next consider two cases. In the first case, we do not include the $G_x$ terms and consider the terms with $G_{1,1}^o$ and $G_{2,2}^o$. By further simplifying these terms, we obtain a compact and physically meaningful expression of transmission
\begin{align}
S_{21}^{o1} = C^{o1} \frac{ \omega-m}{ (\omega-\omega_c)(\omega-\omega_1) }, \label{S21o1m}
\end{align}
where $C^{o1}$ also denotes a constant, $\omega_1 = (4m+\lambda\omega_c)/(4+\lambda)$ and $\lambda=ll_s\epsilon b^2/c_0^2$.

In the second case, we add the $G_x$ terms on the basis of the first case. The expression of transmission becomes
\begin{align}
S_{21}^{o2} = C^{o2} \frac{\omega-m'}{(\omega-\omega'_c)(\omega-\omega'_1)}, \label{S21o2m}
\end{align}
where $\omega'_c=\omega_c-i(1-R)c_0/2l$, $\omega'_1=\omega_1-i4\xi(1-R)[8l(8l/c_0+\xi)/c_0]$, $m'=m-i\Gamma\omega$, and $\xi=2l^2l_s\epsilon b^2/c_0^3$.




\section{III. Results and discussions}

\subsection{IIIA. Numerical results}

Figure~\ref{fig3} shows the tranmission $|S_{21}|$ spectrum obtained by numerically calculating Eq. (\ref{S21m}). One can see three cavity resonance modes, i.e. two odd modes at 510 and 570 GHz, and one even mode at 540 GHz. Due to the strong coupling between the axion mode and even mode of a cavity, a pronounced anticrossing gap occurs in the transmission spectrum. Moreover, as the reflectivity $R$ is increased from 0.99 (Fig.~\ref{fig3}(a)) to 0.999 (Fig.~\ref{fig3}(c)), the anticrossing gap of even mode remains almost unchanged.

For the odd mode, the spectrum is quite different from those of even mode. First, the coupling strength between axion mode and odd mode is very small. As seen in Eq. (\ref{DAF2m}), the coupling is related to the amplitude of electric field $\mathbf{E}$. The electric field at the DAI is minimum and maximum for odd mode and even mode respectively, resulting in a weak coupling between odd mode and axion mode. Second, as shown in the insets of Fig.~\ref{fig3}, our calculations surprisingly present an LA for odd mode, instead of an LR of even mode. This indicates a different coupling mechanism of odd mode from that of even mode. Third, one can see that, when increasing $R$ from 0.99 to 0.999, the size of LA gap decreases.

\subsection{IIIB. Analytical results}

In order to reveal the origin of such an unexpected LA, we study the analytical expressions of $S_{21}$ for even and odd modes given in Sec. IIE. Before doing so, we first compare the numerical and analytical results in order to verify the validity of the derived analytical expressions. Figure~\ref{fig4} shows the transmission $|S_{21}|$ spectrum of a 0.1$\mu m$-thick DAI for the odd mode. One can see a good agreement between numerical results and analytical expression. For both values of reflectivity $R$=0.99 and $R$=0.999, the analytical formula Eq. (\ref{S21o2m}) reproduces the numerical results very well. Therefore, the analytical method and resulting formulae given in Sec. IIE are correct and reliable.

For the even mode, the transmission $|S_{21}|$ spectrum obtained from analytical expression Eq. (\ref{S21em}) is shown in Fig.~\ref{fig5}(a). The LR with upper and lower polariton branches (Fig.~\ref{fig5}(b)) is clearly seen. By comparing the numerical results in Fig.~\ref{fig3}(a) and analytical results in Fig.~\ref{fig5}(a), one can find that the numerical result is well described by the analytical expression Eq. (\ref{S21em}). In deriving this analytical formula, we consider only the zeroth-order and linear-order terms of type-I interaction. This treatment is also justified since the linear term is the largest among all $G_n$ terms as seen in Fig.~\ref{fig5}(c). The agreement between numerical and analytical results implies that the coupling between axion mode and even-order photon mode is of the linear nature. Such a linear coupling and the expression of $S_{21}$ have been obtained for many other coupled systems, e.g. photon-phonon, photon-magnon and photon-exciton coupling, etc~\cite{cohentannoudji2004book,haroche2013rmp,savage1989prl}.

But for the odd mode, the above picture is completely changed. By checking the expansion of the denominator $G$, we found that the first- and second-order terms of type-I interaction, which dominate in the case of even mode, are vanishing (Fig. \ref{fig5}(f)). So, the dominant terms are the high-order type-II terms and type-I terms. As $l_s$ is small, the latter terms become very small and thus the high-order type-II interaction mainly contributes. In Sec. IIE, we derive two formulae by including and excluding the $G_x$ terms. The $G_x$ terms are related to the reflectivity $R$ and thus give rise to the effect of coupling on the line-width of modes. We first discuss the spectrum and thus neglect the $G_x$ terms. The analytical results given in Fig.~\ref{fig5}(d) show that there appears a new high-order mode with frequency $\omega_1$. Since this high-order mode has a smaller slope than the axion mode, it bends to the cavity resonance with an attractive characteristic. Moreover, the high-order mode $\omega_1$ reduces to the axion mode at resonance ($m=\omega_c$), reproducing $S_{21}=C^{o}/(\omega-\omega_c)$ shown in Fig.~\ref{fig5}(e). Therefore, the LA under this study is of both gapless and high-order nature, which is the central result of this work.

\subsection{IIIC. Dissipation effect}

The reflectivity $R$ indicates the decay rate (dissipation) of photons inside a cavity. A smaller $R$ represents a larger photon leakage (dissipation). In Eqs. (\ref{S21em}), (\ref{S21o1m}) and Fig.~\ref{fig5}, we assume the reflectivity $R$=1, implying no leakage of intracavity photons, and we obtained a gapless LA. Here, we further consider the dissipation with $R<1$. Based on te numerical method, we calculate the $S_{21}$ spectrum of the odd mode with $R=0.99$ shown in Fig.~\ref{fig6}(a) and find a gap at resonance, in contrast to the gapless LA in Fig.~\ref{fig5}(d).  Furthermore, a larger dissipation (a smaller $R$) results in a wider gap, shown in Fig.~\ref{fig6}(b). To understand the behavior in Fig.~\ref{fig6}, we deduce Eq. (\ref{S21o2m}) by including the $G_x$ terms in the derivation. Interestingly, the new formula is of the same form as Eq. (\ref{S21o1m}). The only difference is that the dissipation and line-width of coupled modes are renormalized. The results clearly give the correlation between dissipation rate and reflectivity $R$. The new formula satisfactorily reproduces the gapped LA, indicating that the gap originates from the dissipation effect. Further analysis (Fig.~\ref{fig6}(c)) shows that the minimum of the function $(\omega - m')$ at $m$ (see F2) plays an important role in producing the gap. Therefore, the dissipation is not the key to generating the high-order LA, but induces a gap in the LA.

\subsection{IIID. Coupled-wave picture}
To better understand the high-order LA, we present a model in which the photon and axion modes are simplified by waves. Since the photon is restricted inside the cavity, the cavity photon is described by a standing wave. For the axion mode, as pointed out in Sec. IIB, it is of uniform distribution inside the DAI. Therefore, the axion mode is modeled as a traveling wave. The standing wave, expressed by $y_1 \propto \cos(kx) \cos(\omega t)$ with a wave vector $k=\frac{\omega}{v_s}$, sets up in the region $\left[ 0 \sim L \right]$. $v_s$ is the speed of wave and the resonance frequency $\omega_s=n\pi\frac{v_s}{L}$. The traveling wave, $e.g.$ $y_2 \propto \cos(kx - \omega t)$, propagates in the region $\left[ \frac{L-l_s}{2}\sim\frac{L+l_s}{2} \right]$. For simplicity, we consider that both waves have the same frequency. The coupling energy of two waves takes the same form of $E^{st} = \int_{\frac{L-l_s}{2}}^{\frac{L+l_s}{2}} dx y_1 \cdot y_2$ as that in $S_{\text{topo}}$. To gain clearer physical picture, we perform the Taylor expansion around $\omega=\omega_s$ and $l_s=0$ for the coupling energy. Up to the third order, the coupling energy is written as
\begin{equation}
E^{e} \propto \left[ l_s - \frac{l_s}{4} (\delta\omega)^2 \right] a(t) + \left[ \frac{l_s}{2} \delta\omega  \right] b(t), \label{modelE}
\end{equation}
for even mode and
\begin{equation}
E^{o} \propto  \left[ \frac{l_s}{4} (\delta\omega)^2 \right] a(t) - \left[ \frac{l_s}{2} \delta\omega  \right] b(t) , \label{modelO}
\end{equation}
for odd mode. Here, $\delta\omega=(\omega-\omega_s)\frac{L}{v_s}$, $a(t)=\cos^2(\omega t)$ and $b(t)=\cos(\omega t)\sin(\omega t)$. At $\omega=\omega_s$, the coupling energy is nonzero for even mode($n$ even) and thus the coupling induces frequency shift and mode hybridization. But for odd mode($n$ odd), the coupling energy is zero, explaining the gapless feature of high-order LA in cavity axion polariton, $i.e.$ $\omega=\omega_c=m$ at resonance. To the lowest order in Eq. (\ref{modelE}) and (\ref{modelO}), the linear term ($\sim l_s$) dominates for even mode while the high-order term ($\sim l_s \delta\omega$) dominates for odd mode. Therefore, the high-order interaction plays a key role in the coupling between a traveling wave and an odd-order standing wave.



\subsection{IIIE. Discussions}
The LA found in this work is essentially different from the conventional LA. First of all, the LA here is attributed to a high-order interaction, but the conventional LA originates from the linear-order coupling of two modes~\cite{metelmann2015prx,bernier2018pra,harder2018prl,yu2019prl,tserkovnyak2005rmp,bhoi2019prb,grigoryan2018prb,boventer2020prr}. Second, the dissipation is not the essence for the LA in this work and it merely tunes the LA gap. Third, the axion mode under this study arises from longitudinal magnetization fluctuation, instead of transverse magnetization fluctuation in cavity magnon polariton~\cite{soykal2010prl,huebl2013prl,tabuchi2014prl,goryachev2014prapplied,zhang2014prl,bai2015prl,wang2018prl,cao2015prb}.

To demonstrate the LA here is controllable, we vary the coupling strength $b$ and the thickness $l_s$ of the DAI film. As shown in Fig.~\ref{fig1}(e), the topologically trivial DAI possesses a negligible axion-photon coupling strength and thus does not couple with cavity resonance (Fig.~\ref{fig7}(a)). Around the topological transition, the coupling strength quickly increases, resulting in a large gap of LA (Fig.~\ref{fig7}(b) and \ref{fig7}(c)). This trend of increasing gap can also be obtained through increasing the thickness $l_s$. But, extra high-order modes may appear. As for a large $l_s$ case, the type-II interaction does not overwhelm the type-I interaction that is absent for a small $l_s$ case. Figure~\ref{fig8} shows the transmission $| S_{21} |$ spectrum as the reflectivity $R$ increases. One can see that the spectrum changes from an LA at $R$=0.99 and 0.999 to an LR at $R$=0.9999. As explained in Sec. IIIB and IIIC, the LA arises from the high-order type-II interaction and the dissipation effect. But as $R$ increases and approaches unity, the dissipation effect becomes weaker and weaker. In such a case, the higher-order term, e.g. the third-order term, of the type-I interaction starts to contribute and dominates to produces an LR. The high-order terms imply more new modes as shown in Fig.~\ref{fig7}(e) and \ref{fig7}(f). The competition between type-I and type-II interaction results in the coexistence of the LR and LA. Therefore, it is expected to observe the LA, LR and high-order modes by tuning the topological phases ($M$) and the DAI thickness $l_s$.

To make possible experimental measurements, a high-frequency cavity and a high-quality DAI material are important. Recently, Scalari et al. reported the fabrication of a $\mu m$-size electronic meta-material cavity with resonance frequency up to 4 THz~\cite{scalari2012science}, which well covers the frequency ($\sim 0.5$ THz) of DAIs in this work. The ultra-strong coupling with a low dissipation rate has also been achieved, implying the potential application in cavity quantum electrodynamics. Further, thanks to the fast-growing field of MnBi$_2$Te$_4$-based magnetic topological materials and state-of-art experimental techniques~\cite{gong2019cpl,zhang2019mbt,li2019sa,otrokov2019nature,rienks2019nat,deng2020,liu2020robust,chen2019intrinsic,klimovskikh2020npj,yan2019prm,hao2019prx,chen2019prx,zeugner2019,wu2019sa,vidal2019prx,hu2020nc,he2020}, it is promising that candidate materials hosting a large topological DAF, such as MnBi$_2$Te$_4$ films, Mn$_2$Bi$_2$Te$_5$ and MnBi$_2$Te$_4$/Bi$_2$Te$_3$ superlattice proposed by some of the authors~\cite{zhu2020,zhang2020cpl,wang2020dynamical}, can be used to study the high-order LA of cavity axion polariton.

\begin{acknowledgements}
\textbf{Acknowledgements.} This work is supported by the Fundamental Research Funds for the Central Universities (Grant No. 020414380185), the Natural Science Foundation of China (Grants No. 61974067, No. 12074181, No. 12074181, No. 12174158 and No. 11834006), Natural Science Foundation of Jiangsu Province (No. BK20200007) and the Fok Ying-Tong Education Foundation of China (Grant No. 161006).
\end{acknowledgements}

\section{APPENDIX A: Derivation of dynamical equations}

\setcounter{equation}{0}
\renewcommand\theequation{A\arabic{equation}}

Here we show the procedure of deriving the dynamical equations of axion and photon modes. Based on Eqs. (\ref{Saction}) and the Euler-Lagrangian equation, we can obtain
\begin{eqnarray}
 \nabla \cdot \mathbf{D} &=&4 \pi \rho-\frac{\alpha}{\pi}(\nabla \delta \theta \cdot \mathbf{B}) \label{maxwell1}\\
 \nabla \times \mathbf{H} &=&\frac{1}{c_0} \frac{\partial \mathbf{D}}{\partial t}+\frac{4 \pi}{c_0} \mathbf{j}_0 \notag \\
 &+&\frac{\alpha}{\pi}\left((\nabla \delta \theta \times \mathbf{E})+\frac{\mathbf{B}}{c_0}\left(\partial_{t} \delta \theta\right) \right) \label{maxwell2} \\
 \nabla \times \mathbf{E} &=&-\frac{1}{c_0} \frac{\partial \mathbf{B}}{\partial t} \label{maxwell3} \\
 \nabla \cdot \mathbf{B} &=&0 \label{maxwell4} \\
 \frac{\alpha \mathbf{E} \cdot \mathbf{B}}{8 \pi^{2} g^{2} J} &=&\frac{\partial^{2}}{\partial t^{2}} \delta \theta-v^{2} \nabla^{2} \delta \theta+m^{2} \delta \theta, \label{maxwell5}
\end{eqnarray}
where $\rho$ and $\mathbf{j}_0$ are the charge density and current density. Equations (\ref{maxwell1})$\sim$(\ref{maxwell4}) correspond to the modified Maxwell's equations, and Eq. (\ref{maxwell5}) describes the equation of motion for axion field. In this work, we have $\mathbf{j}_0=0$ for antiferromagnetic insulator. Moreover, we consider the long-wavelength axion mode for which the axion field is uniform inside the DAI. Therefore, the dispersion of the axion mode is not taken into account and the term of $\nabla^{2} \delta \theta$ in Eq. (\ref{maxwell5}) is ignored.

Substituting the relation $\mathbf{B}=\mu_r \mathbf{H}$ into Eq. (\ref{maxwell2}) and adding $\partial_{t}$ to both sides of Eq. (\ref{maxwell2}), we obtain
\begin{equation}
 \partial_{t}(\nabla \times \frac{\mathbf{B}}{\mu_r} ) =\frac{\epsilon}{c_0} \frac{\partial^{2} \mathbf{E} }{\partial t^{2}}+\frac{\alpha}{\pi}\partial_{t}\left((\nabla \delta \theta \times \mathbf{E})+\frac{\mathbf{B}}{c_0}\left(\partial_{t} \delta \theta\right) \right).
\end{equation}

Using Eq. (\ref{maxwell3}) and the relation
\begin{equation}
\nabla\times(\nabla\times \mathbf{E})=\nabla(\nabla\cdot \mathbf{E})-\nabla^{2} \mathbf{E},
\end{equation}
we obtain
\begin{equation}
\frac{\partial^2 \mathbf{E}}{\partial t^2} - c^2 \bigtriangledown^2 \mathbf{E} + \frac{\alpha }{\pi \epsilon} \frac{\partial^2 \delta \theta}{\partial t^2} \mathbf{B} = 0.
\end{equation}

In the derivation, the terms which couple $\mathbf{E}$ and $\mathbf{B}$ to $\delta\theta$ are kept to the linear order only.

As for the DAF, Eq. (\ref{maxwell5}) is rewritten as
\begin{equation}
\frac{\partial^2 \delta \theta}{\partial t^2} + m^2 \delta \theta + \Gamma \frac{\partial \delta \theta}{\partial t} - \frac{\alpha }{8\pi^2 g^2 J} \mathbf{B} \cdot \mathbf{E} = 0,
\end{equation}
where the intrinsic damping $\Gamma$ of axion mode is added.

Moreover, the magnetic induction $\mathbf{B}$ consists of both magnetic component of electromagnetic wave and static magnetic field $\mathbf{B}_0$. Since the former is in general four orders of magnitude smaller than the latter, only $\mathbf{B}_0$ is taken into account in these two equations. Hence, the above two equations can be rewritten as
\begin{eqnarray}
\frac{\partial^2 \mathbf{E}}{\partial t^2} - c^2 \bigtriangledown^2 \mathbf{E} + \frac{\alpha}{\pi \epsilon} \frac{\partial^2 \delta \theta}{\partial t^2} \mathbf{B}_0 = 0, \\
\frac{\partial^2 \delta \theta}{\partial t^2} + m^2 \delta \theta + \Gamma \frac{\partial \delta \theta}{\partial t} - \frac{\alpha }{8\pi^2 g^2 J} \mathbf{B}_0 \cdot \mathbf{E} = 0
\end{eqnarray}

\section{APPENDIX B: Derivation of transfer matrix and scattering matrix}

\setcounter{equation}{0}
\renewcommand\theequation{B\arabic{equation}}

Inside a cavity, both left-going and right-going propagating states exist. So, the field is expressed by a linear combination of two states, i.e.
\begin{align}
H &= ( H^+ e^{-jky} + H^- e^{jky} ), \label{hz} \\
E &= Z( H^+ e^{-jky} - H^- e^{jky} ).
\end{align}

Next, we consider the boundary condition of electromagnetic field at the left ($y$=0) and right ($y$=$l_s$) surfaces of a DAI,
\begin{equation}
\left\{
\begin{aligned}
H^0 &= H^+ + H^-  \\
E^0 &= Z(H^+ - H^-) \\
H^{l_s} &= H^+ e^{-jkl_s} + H^- e^{jkl_s} \\
E^{l_s} &= Z(H^+ e^{-jkl_s} - H^- e^{jkl_s}).
\end{aligned}
\right.
\end{equation}

By eliminating $H^{\pm}$, one can obtain
\begin{equation}
\left( \begin{array}{c}
           E^{l_s} \\
           H^{l_s} \\
\end{array} \right) =
\left( \begin{array}{cc}
           \cos(kl_s) \ \ \ \ \ \ -jZ\sin(kl_s) \\
           -\frac{j}{Z}\sin(kl_s) \ \ \ \ \ \ \cos(kl_s) \\
\end{array} \right)
\left( \begin{array}{c}
           E^0 \\
           H^0 \\
\end{array} \right). \label{transfer1}
\end{equation}

The matrix in Eq. (\ref{transfer1}) is the transfer matrix of the DAI. The transfer matrix of the air region is obtained by replacing the quantities $k$, $Z$ and $l_s$ with those in the air region, i.e. $k_0$, $Z_0$ and $l$. Since the cavity contains one DAI and two air regions, there are three transfer matrices. The total transfer matrix is the product of three matrices from left to right and is written as
\begin{equation}
T =
\left( \begin{array}{cc}
           T_A \ \ \ \ \ \ T_B \\
           T_C \ \ \ \ \ \ T_D \\
\end{array} \right).
\label{transferT}
\end{equation}
where $T_{A}\sim T_D$ are the elements of the transfer matrix.

The transfer matrix describes the relation between the fields at input and output ports. In order to calculate the transmission $S_{21}$, we need the amplitudes of incident wave at input port and those of transmitted wave at output port which are connected by~\cite{pozar2011book,yao2015prb}
\begin{equation}
\left( \begin{array}{c}
           E_A^- \\
           E_B^- \\
\end{array} \right) =
\left( \begin{array}{cc}
           S_{AA} \ \ \ \ \ \ S_{AB} \\
           S_{BA} \ \ \ \ \ \ S_{BB} \\
\end{array} \right)
\left( \begin{array}{c}
           E_A^+ \\
           E_B^+ \\
\end{array} \right), \label{SC1}
\end{equation}
where $-(+)$ represents the direction from the cavity (outside) to outside(cavity).

The scattering matrix is derived and written as
\begin{widetext}
\begin{eqnarray}
\left( \begin{array}{cc}
           S_{AA} \ \ \ \ S_{AB} \\
           S_{BA} \ \ \ \ S_{BB} \\
\end{array} \right)  &=& \left( \begin{array}{cc}
          \frac{T_A+T_B/Z_0-T_CZ_0-T_D}{T_A+T_B/Z_0+T_CZ_0+T_D} \ \ \ \ \ \ \frac{2(T_AT_D-T_BT_C)}{T_A+T_B/Z_0+T_CZ_0+T_D} \\
          \frac{2}{T_A+T_B/Z_0+T_CZ_0+T_D} \ \ \ \ \ \ \frac{-T_A+T_B/Z_0-T_CZ_0+T_D}{T_A+T_B/Z_0+T_CZ_0+T_D} \\
\end{array} \right) \notag \\
            &=& e^{2jk_0l} \left( \begin{array}{cc}
           \frac{-\sin(kl_s)(Z_0^2-Z^2)}{(Z_0^2+Z^2)\sin(kl_s)+2jZ_0Z \cos(kl_s)} &  \frac{2jZ_0Z}{(Z_0^2+Z^2)\sin(kl_s)+2jZ_0Z \cos(kl_s)} \\
           \frac{2jZ_0Z}{(Z_0^2+Z^2)\sin(kl_s)+2jZ_0Z \cos(kl_s)} &  \frac{-\sin(kl_s)(Z_0^2-Z^2)}{(Z_0^2+Z^2)\sin(kl_s)+2jZ_0Z \cos(kl_s)} \\
\end{array} \right). \label{Smatrix2append}
\end{eqnarray}
\end{widetext}

Finally, the transmission is written as
\begin{align}
S_{21} = \frac{ (1-R^2) S_{BA}}{(1-RS_{AA})(1-RS_{BB})- R^2S_{AB}S_{BA}}. \label{S21}
\end{align}

Substituting the elements of scattering matrix Eq. (\ref{Smatrix2append}) into Eq. (\ref{S21}), we obtain an analytical expression
\begin{align}
S_{21} = \frac{F}{G}, \label{s21b}
\end{align}
with the numerator
\begin{align}
F = \frac{ 2j e^{-2jk_0l} } {Z_0Z } \left[ (Z_0^2+Z^2)\sin(kl_s)+2jZ_0Z \cos(kl_s) \right]  \label{num}
\end{align}
and the denominator
\begin{align}
G &= 4 - 4(1-R^2)- 4 \cos^2(kl_s)e^{-4jk_0l} \notag  \\
            &+ 4j \frac{Z_0}{Z} \sin(kl_s)\cos(kl_s) (1+e^{-2jk_0l}) e^{-2jk_0l} \notag \\
            &- 4j \frac{Z}{Z_0} \sin(kl_s)\cos(kl_s) (1-e^{-2jk_0l}) e^{-2jk_0l} \notag \\
            &+ \left[ \frac{Z_0}{Z} ( e^{-2jk_0l} + 1) \sin(kl_s) + \frac{Z}{Z_0} ( e^{-2jk_0l} - 1) \sin(kl_s) \right]^2  \notag \\
            &- 2(1-R) e^{-2jk_0l} \left[ \frac{Z_0}{Z} - \frac{Z}{Z_0} \right] \left[ \frac{Z_0}{Z} + \frac{Z}{Z_0} \right] \sin^2(kl_s) \notag \\
            &- 4j(1-R) e^{-2jk_0l} \left[ \frac{Z_0}{Z} - \frac{Z}{Z_0} \right] \sin(kl_s)\cos(kl_s) \notag \\
            &- (1-R^2) \left[ \frac{Z_0}{Z} - \frac{Z}{Z_0} \right]^2 \sin^2(kl_s). \label{denom}
\end{align}

\section{APPENDIX C: Expansion of the denominator G}

\setcounter{equation}{0}
\renewcommand\theequation{C\arabic{equation}}

\textbf{Even mode.} The denominator is written as $G^e = \sum\limits_{n_1=0}^{+\infty} G_{n_1}^e $ with $G_{n_1}^e = \sum\limits_{n_2=0}^{+\infty} G_{n_1,n_2}^e (\Delta \omega_C)^{n_2} (\frac{l_s}{c_0} \omega)^{n_1}$. Here, we present some low-order terms of $G_{n_1}^e$

\begin{align}
(i) G_0^e&=-8 \Delta \omega_C + \cdots.  \label{even-0}
\end{align}

\begin{align}
(ii) G_1^e&= 8j ( \frac{l_s}{c_0} \omega ) (\epsilon \mu - 1) \notag \\
&+ 12j ( \frac{l_s}{c_0} \omega ) (\epsilon \mu - 1) \Delta \omega_C,   \label{even-1}
\end{align}

\begin{align}
(iii) G_2^e&= 4 ( \frac{l_s}{c_0} \omega )^2  (\epsilon \mu - 1)^2 \notag \\
&+ 4 ( \frac{l_s}{c_0} \omega )^2  (\epsilon \mu - 1)^2 \Delta \omega_C,  \label{even-2}
\end{align}

\begin{align}
(iv) G_3^e&= -\frac{4}{3} j ( \frac{l_s}{c_0} \omega )^3 (\epsilon \mu - 1)(\epsilon \mu - \frac{1}{2}),  \notag \\
&- 2j ( \frac{l_s}{c_0} \omega )^3 (\epsilon \mu - 1)(\epsilon \mu - \frac{2}{3}) \Delta \omega_C,   \label{even-3}
\end{align}

\begin{align}
(v) G_4^e&=- \frac{1}{3} ( \frac{l_s}{c_0} \omega )^4  (\epsilon \mu - 1) (\frac{4}{3} \epsilon \mu + 1) (\epsilon \mu + 1), \notag \\
&- ( \frac{l_s}{c_0} \omega )^4  (\epsilon \mu - 1) (\frac{4}{3} \epsilon \mu - \frac{5}{3}) (\epsilon \mu - 1) \Delta \omega_C,  \label{even-4}
\end{align}

\textbf{Odd mode.} The denominator is written as $G^o = G_x + \sum\limits_{n_1=0}^{+\infty} G_{n_1}^o $ with $G_{n_1}^o = \sum\limits_{n_2=0}^{+\infty} G_{n_1,n_2}^o (\Delta \omega_C)^{n_2} (\frac{l_s}{c_0} \omega)^{n_1}$. Here, we present some low-order terms of $G_{n_1}^o$

\begin{align}
(i) G_{0}^o&=-8 \Delta \omega_C + \cdots.  \label{odd-0}
\end{align}

\begin{align}
(ii) G_{1}^o&= 4j ( \frac{l_s}{c_0} \omega ) (\epsilon \mu - 1) \Delta \omega_C,  \notag \\
&+ 6j ( \frac{l_s}{c_0} \omega ) (\epsilon \mu - 1) \Delta \omega_C^2,  \notag \\
&+ \frac{14}{3}j ( \frac{l_s}{c_0} \omega ) (\epsilon \mu - 1) \Delta \omega_C^3,  \notag \\
&+ \frac{5}{2}j  ( \frac{l_s}{c_0} \omega ) (\epsilon \mu - 1) \Delta \omega_C^4,  \label{odd-1}
\end{align}

\begin{align}
(iii) G_{2}^o&= ( \frac{l_s}{c_0} \omega )^2  (\epsilon \mu - 1)^2 \Delta \omega_C^2,  \notag \\
&+ ( \frac{l_s}{c_0} \omega )^2 (\epsilon \mu - 1)^2 \Delta \omega_C^3,  \notag \\
&+ \frac{7}{12} ( \frac{l_s}{c_0} \omega )^2 (\epsilon \mu - 1)^2 \Delta \omega_C^4,   \label{odd-2}
\end{align}

\begin{align}
(iv) G_{3}^o&= \frac{2}{3} j ( \frac{l_s}{c_0} \omega )^3 (\epsilon \mu - 1) ,  \notag \\
&- \frac{2}{3}j ( \frac{l_s}{c_0} \omega )^3 (\epsilon \mu - 1)(\epsilon \mu - 2) \Delta \omega_C,  \notag \\
&- j ( \frac{l_s}{c_0} \omega )^3 (\epsilon \mu - 1)(\epsilon \mu - \frac{4}{3}) \Delta \omega_C^2,  \notag \\
&- \frac{1}{9}j ( \frac{l_s}{c_0} \omega )^3 (\epsilon \mu - 1)(7\epsilon \mu - 8) \Delta \omega_C^3,  \notag \\
&- \frac{1}{36}j  ( \frac{l_s}{c_0} \omega )^3 (\epsilon \mu - 1)(15\epsilon \mu - 16) \Delta \omega_C^4,  \label{odd-3}
\end{align}

\begin{align}
(v) G_{4}^o&=  \frac{1}{3} ( \frac{l_s}{c_0} \omega )^4  (\epsilon \mu - 1)^2, \notag \\
&+ ( \frac{l_s}{c_0} \omega )^4  (\epsilon \mu - 1)^2 \Delta \omega_C,  \notag \\
&- \frac{1}{6} ( \frac{l_s}{c_0} \omega )^4 (\epsilon \mu - 1)^2(2\epsilon \mu - 7) \Delta \omega_C^2,  \notag \\
&- \frac{1}{6} ( \frac{l_s}{c_0} \omega )^4 (\epsilon \mu - 1)^2(2\epsilon \mu - 5) \Delta \omega_C^3,   \notag \\
&- \frac{1}{72} ( \frac{l_s}{c_0} \omega )^4 (\epsilon \mu - 1)^2(14\epsilon \mu - 31) \Delta \omega_C^4,   \label{odd-4}
\end{align}

\begin{align}
(vi) G_x &= 4j ( \frac{l_s}{c_0} \omega ) (\epsilon \mu - 1) (1-R) \notag \\
    &+ ( \frac{l_s}{c_0} \omega )^2  (\epsilon \mu - 1)^2 (1-R)^2 \notag \\
    &+ 2( \frac{l_s}{c_0} \omega )^2 (\epsilon \mu - 1)^2 (1-R) \Delta \omega_C. \label{odd-5}
\end{align}



\bibliography{qah-cavity}

\end{document}